# On the Testing of Seismicity Models

## G. Molchan

[1] International Institute of Earthquake Prediction Theory and Mathematical Geophysics, Russian Academy of Science, Profsoyuznaya 84/32, Moscow, Russia.
[2] The Abdus Salam International Centre for Theoretical Physics, SAND Group, Trieste, Italy.

*E-mail* : *molchan@mitp.ru*

**Abstract**

Recently a likelihood-based methodology has been developed by the Collaboratory for the Study of Earthquake Predictability (CSEP) with a view to testing and ranking seismicity models. We analyze this approach from the standpoint of possible applications to hazard analysis. We arrive at the conclusion that model testing can be made more efficient by focusing on some integral characteristics of the seismicity distribution. This is achieved either in the likelihood framework but with economical and physically reasonable coarsening of the phase space or by choosing a suitable measure of closeness between empirical and model seismicity rate in this space.

## 1. Introduction.

The assessment of seismic hazard and risk are based on maps of long-term rate of damaging seismic events. There is a wide diversity of approaches to the making of such maps, which differ in the use of historical and low magnitude seismicity, seismotectonic regionalization, the Gutenberg-Richter law, smoothing techniques, and so on (see, e.g., *Molchan et al., 1997*; *Giardini, 1999* and references therein). For this reason, the initiative of the U.S. branch of the Collaboratory for the Study of Earthquake Predictability (CSEP) is important; its purpose is to develop a statistical methodology for objective testing and ranking of seismicity models (*Field, 2007*). That program has been implemented as the Regional Earthquake Likelihood Models (RELM) project for California (*Schorlemmer et al., 2010*) and now the methodology is in a stage of active analysis and development (see e.g. *Lombardi and Marzocchi, 2010*; *Werner et al., 2010; Rhoades et al., 2011; Zechar et al.,2010*).

Below we examine the RELM methodology from the point of view of possible applications to hazard analysis, i.e., to the testing of long-term seismicity maps. We provide a brief description of basic elements of the methodology with a parallel discussion of its strong and weak points.

## 2. The Seismicity Model

A seismicity map describes the mean rate of target events $\lambda_j = \lambda(\Delta_j)$ in subsets $\Delta_j$ of space $G \times M$. Here $G$ denotes the region and $M$ the magnitude range of target events. The model $\{\lambda_j\}$ is to be tested based on observations $\{\nu(\Delta_j) = \nu_j\}$ in the subsets $\{\Delta_j\}$ for a period $T$. In the CSEP/RELM experiments, the subsets $\Delta_j$ are standard bins of 10 km in linear size and 0.1 in magnitude; the numbers of events in bins, $\{\nu_j\}$, are generally low: 0,1,2, with the total number

$$N = \sum \nu_j \quad (1)$$

being a few tens of events.

The statistical analysis of maps is based on the following assumption: the variables $\{\nu_j\}$ are independent and poissonian, i.e.,

$$P(\nu_j = k) = \lambda_j^k e^{-\lambda_j} / k!. \quad (2)$$

(the $H_0$ hypothesis).

The $H_0$ hypothesis is reasonable for small time intervals. This circumstance is used by the authors of the methodology to test time-dependent forecasts. They consider the $\{\lambda_j\}$ as a functions of time and in the same time extend the $H_0$ hypothesis to the entire phase space $G \times M \times T$ with an arbitrary $T$. The simplest cluster seismicity models like the Epidemic-Type Aftershock Sequence by *Ogata (1998)* show that the vector $\{\lambda_j(t)\}$ is stochastic and depends on the seismicity under study up to time $t$. On the contrary, $H_0$ assumes that the time variables $\{\lambda_j(t)\}$ are independent for different $t$. This contradiction may cause appreciable difficulties in the testing of non-trivial time-dependent forecasts (see more in *Werner and Sornette (2008), Lombardi and Marzocchi (2010)*).

In testing a long-term seismicity model, the $H_0$ hypothesis is reasonable for main shocks only, i.e., the catalog needs to be declustered. This operation is not unique. Consequently, the statistical analysis should be weakly sensitive to the independence property of $\{\nu_j\}$ as much as possible, focusing on important parameters of the $\lambda(\Delta)$ measure.

## 3. Tests

Nearly all goodness-of-fit tests of model $\{\lambda_j\}$ with data $\{\nu_j\}$ suggested by the RELM working group are based on the likelihood approach. The log-probability of $\{\nu_j\}$
under the $H_0$ hypothesis is

$$L = \sum l(\Delta_j), \qquad (3)$$

where $l(\Delta)$ is

$$l(\Delta) = \nu(\Delta) \ln \lambda(\Delta) - \lambda(\Delta) - \ln \nu(\Delta)! \qquad (4)$$

in virtue of (2); by definition, $\nu! = 1$ for $\nu = 0$.

The $L$-statistic (3) depends on the partition $\{\Delta_j\}$ and the model $\{\lambda(\Delta_j)\}$. The partition is the important component of the methodology, because the partition can be used to good advantage in the testing problem. The simplest examples of the partition are related to the following representation of the rate measure:

$$\lambda(\Delta) = \Lambda p(\Delta g) q(\Delta M), \tag{5}$$

where $\Lambda$ is the rate of target events in $G$, $p(\Delta g)$ and $q(\Delta M)$ are normalized distributions of the events over space and over magnitude, respectively.

Taking the case of the trivial partition as represented by the single element $\Delta = G \times M$, we arrive at the statistic $\nu(\Delta) = N$. The $L$-statistic in this case labeled as $L(N)$ is then given by the relation (4); the distributions of $L(N)$ and $N$ depend on the single parameter $\Lambda = \lambda(G \times M)$, i.e., $N$ is sufficient statistic for the analysis of $\Lambda$.

If the partition deals with magnitude only then $\Delta_j = G \times \Delta_j M$ and the $L$-statistic, $L(M)$, is sufficient for the analysis of the parameters $\{\Lambda q(\Delta_j M)\}$. In practice we use this partition to analyse the frequency-magnitude law.

Finally, the space partition is based on $\Delta_j = \Delta_j G \times M$; the corresponding $L$-statistic, $L(G)$, is sufficient for the analysis of the parameters $\{\Lambda p(\Delta_j G)\}$.

The conditional distribution of $\{\nu_j\}$ provided $\sum \nu_j = N$ is the multinomial distribution with parameters $\{p_1,...,p_n; N\}$, $p_j = p(\Delta_j G)$:

$$P(\nu_1 = k_1,...\nu_n = k_n \mid N) = \frac{N!}{k_1!...k_n!} p_1^{k_1}...p_n^{k_n}, \qquad k_1 + ... + k_n = N. \tag{6}$$

The respective $L$-statistic for the conditional distribution of $\{\nu_j\}$ is

$$L(G \mid N) = \sum_{j=1}^{n} \nu_j \ln p_j + \ln N! - \sum_{j=1}^{n} \ln(\nu_j!). \tag{7}$$

The statistics $N$ and $L(G \mid N)$ enable us to perform a separate analysis of the parameters $\Lambda$ and $\{p_j\}$ (see, e.g., *Molchan and Podgaetskaya*, *1973*; *Molchan et al., 1997; Werner et al., 2010*). The necessity for the separate analysis is caused by many things: the small amount of data $N$, catalog declustering, standardization of catalog magnitude, etc. (see, e.g., *Kagan, 2010,* and *Werner et.al, 2011*).

**The significance of the *L*-test**.

The Monte Carlo method can be used to find the distributions of all type $L$ statistics under $H_0$. In the case $L(G \mid N)$ the distribution (6) corresponds to the model of $N$ independent trials with $n$ outcomes and probabilities $\{p_1,...p_n\}$.

The distribution of $L$ can be used to find the observed significance level for an observed $L$-value, $\hat{L}$:

$$\hat{\alpha} = P\{L \leq \hat{L} \mid H_0\}.$$

In the cases $L: L(M), L(G\mid N), L(G)$ the hypothesis $H_0$ is rejected, if $\hat{\alpha}$ is below the nominal significance level $\alpha_0$ (the conventional value $\alpha_0 = 0.05$ is used). In the case of statistic $N$, both small and large values are suspect, so a two–sided test is used: $\min(\hat{\alpha}, 1-\hat{\alpha}) \leq \alpha_0/2$.

This is a standard scheme for testing any hypothesis. The key point for applications in this scheme is the choice of the test statistic.

**4. Why *L*?**

To answer this question, let us discuss some peculiarities of the RELM experiment:

- in general, the number of tested models $\lambda(\Delta)$ for the same territory can be arbitrary. This is naturally due to the existence of different approaches to create such models. Some local change in a test model can be considered as a new test model. The past seismicity may only impose some integral limitation on the $\lambda(\Delta)$;

- the partition $\{\Delta_j, j = 1, ..., n\}$ is usually very detailed, therefore $n$ is large and the numbers of events $\{\nu_j\}$ are small;

- any possible local relations between $\{\lambda_j\}$ are not used. To be specific if the bins $\{\Delta_j\}$ are small we could assume that the $\lambda_j$ are equal within some space structures;

In other words, in the RELM experiment we have to deal with the statistical problem of a large number of degrees of freedom $f$ because usually $f = n$. The advantage of the likelihood method in such conditions is not obvious. *D.R.Cox and D.V.Hinkley (1974)* in their book "*Theoretical Statistics*" tried to formulate some general principles underlying the theory of statistical inference. One of the obstacles that impede the use of likelihood theory is worded as follows: "in considering problems with many parameters one generally focuses on a small number of components, but to do this one needs principles that are outside the "pure" likelihood theory" (Section 2.4.VIII).

In the hazard evaluation case we are interested in the accuracy of the rate measure $\lambda(\Delta)$. However small values $\{v_j\}$ carry little information for this purpose and rejection of the $\{H_0, \lambda\}$-hypothesis does not mean that the model $\lambda(\Delta)$ is unsatisfactory. It is possible that some components of the hypothesis, i.e., independence of $\{v_j\}$ or the Poisson property may be violated.

In the framework of the likelihood approach we possess a good enough tool to focus on the essentials in the rate measure. The tool in question is the partition of the phase space. For purposes of seismic risk analysis the physically reasonable partitions with $\{v_j\}$ that are not small are preferable. The effect of small/large values of $\{v_j\}$ can be observed by examining the $L(G|N)$-statistic for $N \gg 1$.

**4.1. The case of large $\{v_j\}$.**

Using (7) and the following approximation

$$v! \approx (v/e)^v \sqrt{2\pi v}$$

(8)

for large $\{v_j\}$ and $N$, we can represent $L(G|N)$ as follows:

$$-N^{-1} L(G|N) \approx \sum_{j=1}^{n} \hat{p}_j \ln(\hat{p}_j / p_j) := \rho(\hat{P}, P), \qquad (9)$$

where $x \ln x = 0$ for $x = 0$.

Here, $\hat{P} = \{\hat{p}_j = v_j / N\}$ is an empirical analogue of the distribution $P = \{p_j = \lambda_j / \Lambda\}$, while $\rho(P_1, P_2)$ is the well-known Kullback-Leibler entropy distance from $P_1$ to $P_2$ (see, e.g., *Harte and Vere-Jones, 2005*). This distance is non-negative but is not a metric; for example, $\rho(P_1, P_2) \neq \rho(P_2, P_1)$. It is important that $\rho(P_1, P_2) = 0$ if and only if $P_1 = P_2$.

**Consistency of the $L(G|N)$-test.**

By (6), we have $\hat{p}_j \to p_j^t$ as $N$ becomes large; here $\{p_j^t\} = P^t$ is the true distribution. Consequently, $\rho(\hat{P}, P) \to \rho(P^t, P)$ for $N \gg 1$. We can use this fact to conclude that P is the true distribution if $\rho(\hat{P}, P) \to 0$. Indeed, the relation $\rho(P^t, P) = 0$ implies $P = P^t$. In statistical terms it means that the procedure of selecting the correct model based on small values of $\rho(\hat{P}, P)$ or $|N^{-1}L(G|N)|$ is *consistent* (Borovkov, 1984).

We remind this useful notion for the general situation. Suppose we accept the $P$-model when vector $\{v_j\}$ belongs to some set $\Omega_N$. Suppose this rule guarantees that the P-probability to reject the $P$-model is fixed, e.g., is equal to $\alpha_0$. By definition, the rule is consistent if the $P^t$-probability to accept the $P$-model goes to zero as $N \to \infty$. In other words, any consistent test must reject an incorrect model almost surely as $N$ becomes large. This natural property is highly desirable in the selection problem of the correct model.

In our case $\Omega_N = \{\rho(\hat{P}, P) < c_N\}$, where $c_N \to 0$ because the distribution of $\rho(\hat{P}, P)$, provided $P$ is true, is concentrated close to 0 as $N$ increases indefinitely. But then the relation $\rho(\hat{P}, P) < c_N$ is impossible for an incorrect distribution $P$ for large enough $N$ because the $\rho(P^t, P)$ is strictly positive.

**The $\chi^2$-test**.

The notion of the entropy distance from an empirical to the corresponding theoretical distribution becomes quite transparent in the case $N \gg 1$. We have $\hat{p}_j \to p_j^t$, hence

$$\ln \hat{p}_j / p_j^t \approx (\hat{p}_j - p_j^t) / p_j^t. \qquad (10)$$

Using (9), we get

$$\rho(\hat{P}, P^t) \approx \sum_{j=1}^{n} \frac{(\hat{p}_j - p_j^t)^2}{p_j^t}. \qquad (11)$$

The right-hand side of (11) is well known as the chi-square statistic. For large $N$ in the conditions $H_0$ the statistic $N\rho(\hat{P}, P^t)$ is approximately distributed as chi-square with *f=n-1* degrees of freedom, hence the quantity $-L(G|N) \approx N\rho(\hat{P}, P^t)$ is of the order of $n$.

**Splitting of $L(G)$**

Note that

$$L(G) = L(N) + L(G|N). \qquad (12)$$

By (4), one has $L(N) = N\varphi(\Lambda/N)$ where $\varphi(x) = \ln x - x + 1 \leq 0$ and $\max \varphi(x) = \varphi(1) = 0$, i.e., $L(N) > cN$ if $|\Lambda/N - 1| > \rho$. As we have shown, the contribution of $L(G|N)$ in (12) is of order $n$ in the case of the true $P$-model. Hence, the term $L(N)$ will dominate (12), if $\{p_j\}$ is a satisfactory model but $\Lambda$ is not. This circumstance emphasizes the need for separate analysis of $\Lambda$ and $\{p_i\}$. In practice there are other reasons for this: the parameter $\Lambda$ is very unstable because the declustering operation is not unique, and the Poisson distribution of $N$ is questionable (Kagan, 2010).

### 4.2. The case of small $\{v_j\}$

This case with $v_j = 0$ or 1 is typical of the RELM model experiments because of detailed space partition. As a result, one has $L(G|N) = N\xi_N + N\ln|\Delta| + \ln N!$ where

$$\xi_N = \frac{1}{N}\sum_{j=1}^{n} v_j \ln \dot{p}_j, \quad v_j = 0 \text{ or } 1, \quad \dot{p}_j = p_j/|\Delta|,$$

and $|\Delta|$ is the bin volume. A similar representation of the likelihood function was recommended for the testing of seismicity models by *Rhoads et al. (2011)*.

To analyse the situation when $N \gg 1$, the $\{v_j\}$ values and bin size are small, we have to modify the statistic $\xi_N$ as follows:

$$\xi_N = \int \hat{p}(dg) \ln \dot{p}(g) \ . \tag{13}$$

Here $\hat{p}(\Delta) = v(\Delta)/N$ is the empirical distribution of the events in space, and $\dot{p}(g)$ is density of the rate model measure $p(\Delta)$. If the density is constant within the partition elements then both representation of $\xi_N$ are identical.

The random variable $\xi_N$ given $H_0$ is sum of $N$ independent identically distributed random variables $\ln \dot{p}(g_j)$, where $g_j$ are random event locations in the space $G$ with a distribution $X := x(\Delta)$; more exactly, $X = P$ for the $P$-model and $X = P^t$ for the true model. Therefore $\xi_N$ is approximately Gaussian with the following mean $m(X)$ and variance $\sigma^2(X)/N$:

$$m(X) = \int x(dg) \ln \dot{p}(g),$$

$$\sigma^2(X) = \int x(dg) \ln^2 \dot{p}(g) - m^2(X).$$

Thus, the critical zone $\Omega_N$ of size ~95% for acceptance of model $P$ looks as follows:

$$N^{-1}\xi_N > m(P) - 2\sigma(P)/\sqrt{N}. \qquad (14)$$

If $m(P) = m(P^t)$, then $P^t$ probability of $\Omega_N$ is >50%. Indeed, given $P^t$ we have the following representation:

$$\xi_N = m(P^t) + \xi\sigma(P^t)/\sqrt{N},$$

where $\xi$ is approximately a standard Gaussian variable. Substituting this relation in (14) and taking the equality $m(P) = m(P^t)$ into account, we have

$$P^t(\Omega_N) = P(\xi > -2\sigma(P)/\sigma(P^t)) > P(\xi > 0) \approx 0.5.$$

Obviously, the relation $m(P) = m(P^t)$ does not yield the equality $P = P^t$, because this relation holds for any pair $p(\Delta)$ and $p^t(\Delta) = p(\Delta) + \varepsilon\psi(\Delta)$, where $\psi$ is orthogonal to the following functions: $\varphi_1(g) = 1$ and $\varphi_2(g) = \ln \dot{p}(g)$, i.e., $\int \psi(dg) = 0$, $\int \psi(dg) \ln \dot{p}(g) = 0$.

Hence, the likelihood approach in the case of small $\{v_j\}$ is not consistent because $P^t(\Omega_N) > 0.5$ for large enough $N$.

## 5. Comparison of the models

*Schorlemmer et al. (2007)* tried to rank the tested models using pairwise comparison of the hypotheses $H^{(i)} = \{H_0, \lambda^{(i)}(\Delta)\}$. To test $H^{(1)}$ vs. $H^{(2)}$, the statistic $R = L^{(1)} - L^{(2)}$ is used ($L^{(i)}$ is the $L$-statistic for the $H^{(i)}$ hypothesis). A small observed value of $R$ under $H^{(1)}$, i.e.,

$$P\{R \leq \hat{R} \mid H^{(1)}\} \leq \alpha_0,$$

is treated as evidence for the $H^{(2)}$ model. In classical approach the $R = L^{(1)} - L^{(2)}$ test operates with unique reference model (see the Neyman-Pearson Lemma, *Cox and Hinkley, 1974*). On the contrary, in the CSEP/RELM approach any of tested models is considered as a reference, i.e., the direction in the test procedure is lost. Therefore interpretation of the testing results in such case might be difficult. We now consider this point.

Proceeding as above, we shall restrict ourselves to the parameters $\{p_1,...p_n\}$ alone. The analogue of $R$ is then the statistic $R(G\,|\,N) = L^{(1)}(G\,|\,N) - L^{(2)}(G\,|\,N)$. In virtue of (7) one has

$$N^{-1}R(G\,|\,N) = \sum_{j=1}^{n} \hat{p}_j \ln p_j^{(1)}/p_j^{(2)} = \Sigma \hat{p}_j \ln \hat{p}_j/p_j^{(2)} - \Sigma \hat{p}_j \ln \hat{p}_j/p_j^{(1)},$$

or

$$N^{-1}R(G\,|\,N) = \rho(\hat{P},P^{(2)}) - \rho(\hat{P},P^{(1)}), \qquad (15)$$

As above, $\hat{P} = \{\hat{p}_j\}$ is the empirical distribution. In contrast to the approximate relation between $L(G\,|\,N)$ and the entropy distance, the relation (15) is exact. The fact $R(G\,|\,N) > 0$ means that the empirical distribution $\hat{P}$ is closer to $P^{(1)}$ than to $P^{(2)}$.

The linearity of $R(G\,|\,N)$ in the data $\{v_j\}$ makes it easier to estimate the distribution of this statistic (*Rhoades et al, 2011*), because for large $N$ the quantity $N^{-1}R(G\,|\,N)$ is approximately normal and under $H^{(1)}$ has the mean

$$m = \sum_{j=1}^{n} p_j^{(1)} \ln p_j^{(1)}/p_j^{(2)} = \rho(P^{(1)},P^{(2)}) \qquad (16)$$

and the variance $\sigma^2/N$ where

$$\sigma^2 = (\Sigma p_j^{(1)} \ln^2 p_j^{(1)}/p_j^{(2)} - m^2). \qquad (17)$$

The following example demonstrates the difficulties arising in the $R(G\,|\,N)$ test interpretations.

**Example.** Let us consider a model $P^{(1)} = (p_1, p_2...p_{n-1}, p_n)$ and the dual model $P^{(2)} = (p_n, p_2...p_{n-1}, p_1)$, in other words, $P^{(1)}$ and $P^{(2)}$ are only different in two bins $\Delta_1$ and $\Delta_n$. It is convenient to use the following representation:

$$p_1^{(1)} = a + \delta, \quad p_n^{(1)} = a - \delta \quad \text{and} \quad p_1^{(2)} = a - \delta, \quad p_n^{(2)} = a + \delta,$$

where $a > \delta$ and $a + \delta < 1$. Then

$$N^{-1}R(G\,|\,N) = \frac{v_1 - v_n}{N} \ln \frac{a+\delta}{a-\delta}. \qquad (18)$$

Considering the case of large $\nu_1, \nu_n, N$, we can have $R(G|N) > 0$ with $P^{(1)}$-probability $\geq 0.95$ for a suitable parameters $(a, \delta, N)$. Indeed, according to (16, 17), we have with probability $>0.95$

$$N^{-1} R(G|N) \geq m - 2\sigma / \sqrt{N},$$

where

$$m = 2\delta \ln \frac{a+\delta}{a-\delta} \quad \text{and} \quad \sigma = \sqrt{2a - 4\delta^2} \ln \frac{a+\delta}{a-\delta}.$$

The threshold $m - 2\sigma / \sqrt{N}$ is positive if $\delta^2 > 2a/(N+4)$; for example, if $a = 0.01$, $\delta = 0.009$, and $N = 300$.

Now impose a single requirement on the true model, namely, $p_1^t = p_n^t$. Then one has $P^t(\nu_1 \leq \nu_n) = P^t(\nu_1 \geq \nu_n)$ and therefore the true distribution of the $R$-statistic (18) will be symmetrical. This entails instability of the inferences based on $R(G|N)$.

To show this, we proceed as follows. Suppose we observe $\nu_1 \leq \nu_n$, then we have $R(G|N) \leq 0$. Given $P^{(1)}$, the model $P^{(1)}$ will be rejected in favor of $P^{(2)}$ at confidence level 95%. That means that the relation $\rho(\hat{P}, P^{(2)}) \leq \rho(\hat{P}, P^{(1)})$ is significant under $P^{(1)}$ or briefly $(P^{(2)} \succ P^{(1)} | P^{(1)})$, where '$\succ$' means 'better than'.

Due to duality of distributions $P^{(1)}$ and $P^{(2)}$ we shall have the diametrically opposite inference, namely $(P^{(1)} \succ P^{(2)} | P^{(2)})$, if $\nu_1 \geq \nu_2$. Hence, in the case $\nu_1 = \nu_n$ the two models will reject each other.

The CSEP group encountered this seeming contradictory situation empirically (*Gerstenberger et al., 2009*). *Rhoads et al. (2011)* explained it as follows: "In fact, an $R$-test rejection of model $P^{(1)}$ does not imply anything regarding the superiority of model $P^{(2)}$; it simply indicates that the observed catalogue is inconsistent with model $P^{(1)}$". The last sentence is not entirely accurate. When the $R(G|N)$ test rejects $P^{(1)}$-model relative to a reference model $P^{(2)}$, that means that $P^{(1)}$ has significant local departures from $P^{(2)}$. This does not mean, however, that $P^{(1)}$ can be far from the true distribution in the area where both models are identical because the contribution of this area into the R-test is zero.

**Inconsistency of the *R*-test**.

We can consider $R(G|N)$ as a possible statistic for testing the model provided $P^{(2)}$ is a reference distribution. In such case the $R$-test will not necessarily be consistent, that is, the false model will not always be rejected as N becomes large.

The proof of this statement is the same as in previous section 4.2 The key point here is the following. In the case of large $\{v_j\}$ the random variable $\xi_N = N^{-1}R(G|N)$ given $(H_0, P)$ is approximately Gaussian with mean $m(P) = \rho(P, P^{(2)}) - \rho(P, P^{(1)})$ and variance of the type $\sigma^2(P)/N$ (see (17)). By arguments of section 4.2, the R-test is consistent if the relation $m(P^{(1)}) = m(P^t)$ results in $P^{(1)} = P^t$. However, this is not true.

We can illustrate our assertion using the dual models $P^{(1)}, P^{(2)}$ (but not $P^t$) from our example. One has

$$m(P^t) = (p_1^t - p_n^t)\ln\frac{a+\delta}{a-\delta} \quad \text{and} \quad m(P^{(1)}) = (p_1^{(1)} - p_n^{(1)})\ln\frac{a+\delta}{a-\delta}. \quad (19)$$

These quantities are equal if one has $p_1^t - p_n^t = p_1^{(1)} - p_n^{(1)}$ only while other bin probabilities are arbitrary.

## 6. The area skill score, $A_N$

The area skill score test, $A_N$, was put forward by *Zechar and Jordan (2010a,b)*. The test is based on the following helpful idea. Consider a family of subareas $U_h$ of the region $G$ that increases if $h$ does:

$$U_0 \subset U_h \subset U_{h'} \subset G, \quad h < h' \quad (20)$$

For example, the family may be linked to levels of some positive function in G:

$$U_h = \{g : u(g) \le h\}.$$

The function $u(g)$ may be specified as a prior rate of seismic events within a magnitude range.

Suppose $\hat{F}_N(h)$ gives the relative number of target events that fall in $U_h$ for some test period. Accordingly,

$$F(h) = P(U_h) := \int_{U_h} P(dg) \quad (21)$$

is the probability of a target event falling in $U_h$ under $\{H_0, P\}$. In this way we have reduced the problem of agreement between the data and the model to the classical problem of agreement between an empirical and a theoretical distribution. The freedom of choice for $U_h$ allows one to focus on those elements in the models $\{P^{(i)}\}$ which are important for the investigator and to group the available observations accordingly.

The statistic $A_N$ is defined by *Zechar and Jordan (2010a)* in terms of the $(n,\tau)$ diagram (*Molchan, 1997*), but in the situation we are considering it can be represented as follows:

$$A_N - 1/2 = \int_0^\infty (\hat{F}_N(h) - F(h))dF(h). \tag{22}$$

When $N$ is large, the empirical distribution converges to the true distribution, $F^t(h)$, i.e., (22) $\to 0$ if $F(h) = F^t(h)$. But this is true also for any distribution of the type $F(h) = \phi(F^t(h))$, where $\phi(x)$ is an arbitrary distribution on (0,1) with mean ½. This fact means that the $A_N$-test is not consistent. The formal way to prove the statement is the same as in section 4.2.

Namely, the random variable $\xi_N = A_N - 1/2$ has the following representation in terms of $N$ independent identically distributed variables $\{h_k\}$:

$$\xi_N = 1/2 - N^{-1}\sum F(h_k).$$

The distribution of $h_k$ is $F(h)$ for the case of a $P$ model and $F^t(h)$ for the case of the true model $P^t$. Thus $\xi_N$ is asymptotically Gaussian with mean $m(P) = 0$ for the $P$ case and

$$m(P^t) = 1/2 - \int_0^\infty F(h)dF^t(h) = \int_0^\infty (F^t(h) - F(h))dF(h)$$

for the $P^t$ case while the variance of $\xi_N$ for any model is of order $N^{-1/2}$.

We have already shown that the solution of the equation $m(P) = m(P^t)$ is not unique and is given by the formula $F(h) = \phi(F^t(h))$. Therefore, proceeding as in section 4.2, we can conclude that the $A$-test is not consistent.

The situation can be changed by transforming (22) into a distance as follows:

$$\rho(\hat{F}, F) = \int_0^\infty |\hat{F}_N(h) - F(h)|^\alpha \psi(F(h))dF(h),$$

where $\psi(x)$ is some weight function. The case $\alpha = 2$ with $\psi(x) = 1/x(1-x)$ or $\psi(x) = 1$ gives the well-known nonparametric $\omega^2$-test (*Bolshev and Smirnov, 1983*). The Kolmogorov-Smirnov (K-S) test gives us another example of distance (even a metric):

$$\rho(\hat{F}, F) = \max |\hat{F}(h) - F(h)|.$$

*Rhoades et al., (2011)* successfully applied the K-S test in the framework of the CSEP experiments.

In other words, a reasonable choice of the metric can be a natural alternative to the likelihood method. The non-trivial part of this approach consists in the choice of the sequence of subsets $U_h$ which depend on the goals of research.

## 7. Conclusion

The CSEP experiment deals with testing and ranking of seismicity rate models. In this approach there is no prior limitation on the number of models, all models are equally a priori acceptable, the number of partition elements of phase space, $n$, to group the data is large. Under these conditions the advantage of the likelihood (LH) method that is used as the main tool is not obvious.

We analyzed theoretically the LH method in two particular cases: (1) numbers of events $\{v_j\}$ in space bins are large, which can be of interest for the testing of long-term seismicity maps, and (2) the $\{v_j\}$ are small, which is typical of the CSEP experiments. In the second case, LH method loses a highly desirable property, namely, statistical consistency. In other words, there exist nontrivial models which cannot be classified as wrong by the LH method as the number of observations $N$ becomes large. The same is true regarding the other tests being used under the less stringent limitations on $\{v_j\}$ (the $R$ and the Area Skill Score tests).

The case of small $\{v_j\}$ arises from the detailed partition of the phase space, i.e., when $n$ is large. As a result, an additional undesirable property of the test methodology appears. The testing procedure is based on the rate model and on the assumption of independence of the variables $\{v_j\}$. Selection of the correct rate model is the most important part of the testing while the independence property is usually questionable. The greater $n$ is, the more the independence property affects on the statistical conclusions. Consequently the statistical test analysis should be weakly sensitive to this property as much as possible, focusing on important elements of the rate measure. Practically this is achieved 1) by a physically reasonable coarsening of the phase space, and 2) by choosing a suitable measure of closeness between empirical and model seismicity rate in the space. A formal realization of this idea is presented in section 6 as a generalization of the Area Skill Score test.

**Acknowledgments**

I am grateful to reviewers for careful reading of the paper, useful comments and stimulating discussion.